\documentstyle[preprint,aps]{revtex}

\newcommand{\be}{\begin{eqnarray}}
\newcommand{\ee}{\end{eqnarray}}
\newcommand{\eps}{\varepsilon}
\newcommand{\e}{ {\rm e} }
\newcommand{\la}{\big\langle}
\newcommand{\ra}{\big\rangle}

\renewcommand{\o}{\omega}
\renewcommand{\l}{\lambda}
\renewcommand{\t}{\tilde}
\renewcommand{\jmath}{j}
\renewcommand{\vec}[1]{\mbox{\boldmath$#1$}}

\begin{document}



\draft

\preprint{TPR-96-22}

\title{Photon Spin and the Shape of the Two-Photon Correlation Function}

\author{Claus Slotta and Ulrich Heinz}

\address{Institut f\"ur Theoretische Physik, Universit\"at Regensburg,\\
         D-93040 Regensburg, Germany}

\date{\today}

\maketitle

\begin{abstract}

We use the covariant current formalism to derive the general form of
the 2-photon correlation function for fully chaotic sources. Motivated
by the recent discussion in the literature we concentrate on the effects
from the photon spin on the correlator. We show that for locally
thermalized expanding sources, like those expected to be created in
relativistic heavy ion collisions, the only change relative to 2-pion
interferometry is a statistical factor ${1\over 2}$ for the overall strength
of the correlation which results from the experimental averaging over
the photon spin.

\end{abstract}

\narrowtext


The hot and dense strongly interacting matter created in relativistic
heavy-ion collisions has now for many years been a subject of intense
investigations. While the bulk of the particles emitted in such collisions
are hadrons which due to their late decoupling provide direct information
only on the final state if the collision region (its space-time geometry
as well as its thermal and dynamical properties), photons and dileptons
are emitted directly and thus may serve as a probe also for the early
stages of the reaction zone. The absence of any discernible final state
interactions makes direct photons a particularly clean and desirable
probe for precision experiments like two-particle interferometry to
extract the ``source size''; their immediate usefulness is, however,
severely curtailed by large background contaminations due to
electromagnetic decays of hadrons and hadron resonances after freeze-out.
2-photon interferometry has so far been used successfully only at low
energies ($E_{\rm lab} < 100$ MeV/nucleon) where the $\pi^0$ decay
background can be controlled \cite{M94} but where photon production
results from a complicated interplay of coherent and incoherent production
mechanisms \cite{BKWB96,PPT95}. At ultrarelativistic energies, where
incoherent (``chaotic'') emission processes are expected to dominate
\cite{N86,SK93,TPRW94}, so far no clear direct photon signal has been
observed \cite{WA80}.

Nevertheless several authors have considered two-photon correlations for
high energy heavy ion collisions theoretically \cite{N86,SK93,TPRW94,RF95}.
In this context a controversy has arisen as to how the photon spin should
be correctly accounted for \cite{N86,SK93,RF95}. In the simplest version
\cite{SK93} the correlator is simply multiplied by a statistical factor
${1\over 2}$ to account for the effect of spin averaging in the detector
because only photons with identical spin polarization contribute to
the Bose-Einstein correlations. In Refs.~\cite{N86,RF95} additional
momentum-dependent factors were suggested which were argued to arise
from incomplete overlap of the polarization vectors at non-zero
relative momentum between the photons and which affect also the shape
(not only the normalization) of the correlator. The detailed form of this
momentum-dependent modification differs between Refs.~\cite{N86,RF95}.

In this note we attempt to settle this issue by a careful rederivation
of the two-photon correlation function using the covariant current
formalism \cite{GKW79,CH94}. Our final result supports the simple
normalization prescription used in Refs.~\cite{SK93}. The additional
momentum dependent terms of Refs.~\cite{N86,RF95} are argued to be
artifacts, arising from neglecting the constraints of charge conservation
in the case of Ref.~\cite{N86} and of other very general algebraic
constraints on the tensor structure of the photon emission function in
Ref.~\cite{RF95}.

Our treatment will follow rather closely the formalism developed in
Refs.~\cite{GKW79,CH94}, generalizing it to vector fields. The covariant
single- and two-particle photon distributions are defined by
 \begin{eqnarray}
 \label{P1}
   P_1(\vec{k},\l)
   &=& \o \, \la \hat a^{\dagger}(\vec{k},\l) \, \hat a(\vec{k},\l) \ra \;,
 \\
 \label{P2}
   P_2(\vec{k}_a,\l_a; \vec{k}_b,\l_b)
   &=& \o_a \, \o_b \, \la \hat a^{\dagger}(\vec{k}_a,\l_a) \,
                           \hat a^{\dagger}(\vec{k}_b,\l_b) \,
   \hat a(\vec{k}_b,\l_b) \, \hat a(\vec{k}_a,\l_a) \ra \, .
 \end{eqnarray}
Here $\hat a^{\dagger}(\vec{k},\l)$ creates a photon with momentum
$k = (\o,\vec{k})$, $\o = \vert \vec{k}\vert$, and with polarization $\l$,
and $\la \dots \ra$ denotes the trace over the density matrix of the
photon emitting source. Following \cite{GKW79} we consider this source
as an ensemble of ``elementary'' classical electromagnetic currents
$j_\mu(x)$ and construct the density matrix from eigenstates of the solution
of Maxwell's equations (in Lorentz gauge $\partial\cdot\hat A(x) =0$)
 \be
 \label{Maxwell}
   \Box \hat A_\mu(x) = J_\mu(x) \, .
 \ee
The classical source current $J_\mu(x)$ is taken as a superposition of
elementary currents (wave packets) $j_\mu$, centered at space-time points
$x_n$ in the source and boosted to 4-momenta $p_n$ relative to a global
reference frame, emitting photons with a random initial phase $\phi_n$:
 \be
 \label{currentx}
   J_\mu(x) = \sum_{n=1}^N e^{i\phi_n} \, j^{\{x_n,p_n\}}_\mu (x)
   \equiv  \sum_{n=1}^N e^{i\phi_n} \,
   \Bigl(\bar\Lambda_n j\Bigr)_{\!\mu}\!(x-x_n) \, .
 \ee
$x,x_n,p_n$ are coordinates in the global reference frame. $x_n$ denotes
the space-time position of the center of the elementary source $n$, $p_n
= (E_n,\vec{p}_n)$ its 4-momentum on the global frame. $\Lambda_n \equiv
\Lambda(\vec{p}_n)$ describes the boost by momentum $\vec{p}_n$ from the 
global frame into the rest frame of elementary source $n$, and 
$\bar\Lambda_n \equiv \Lambda^{-1}_n = \Lambda(-\vec{p}_n)$ its inverse.
$j_\mu$ thus denotes the elementary current in its own rest frame, 
and $(\bar\Lambda_n j)_\mu$ is the corresponding 4-vector in the global 
frame\footnote{(Eq.~(\ref{currentx}) corrects a 
notational inaccuracy in Ref.~\cite{CH94} which has, however, no further 
consequences for the calculations presented there.}. All elementary 
currents $j_\mu$ have the same internal structure, i.e. the same functional 
dependence on the relative coordinate $x'=\Lambda_n(x-x_n)$ in their 
own rest frame. Note that the 4-momenta $p_n$ of the elementary sources 
are not on the photon mass-shell; on-shell photon momenta will always 
be labelled by $k$ or $\vec{k}$. The Ansatz (\ref{currentx}) is 
completely general; in particular, it allows for arbitrary $x$--$p$ 
correlations as they exist, for example, in expanding sources.

The current (\ref{currentx}) has the on-shell Fourier transform
 \be
 \label{Jk}
   \t{J}_\mu(\vec{k}) = \int d^4x \,
   e^{i k \cdot x} \, J_\mu(x)
   = \sum_{n=1}^N e^{i\phi_n} \, e^{i k \cdot x_n} \,
     \left(\bar\Lambda_n \t{\jmath}\right)_{\!\mu}\!\!(k) \, ,
 \ee
where
 \be
 \label{jk}
   \left(\bar\Lambda_n \t{\jmath}\right)(k) 
   = \bar\Lambda_n\left(\t{\jmath}(\Lambda_n k)\right)
   = \bar\Lambda_n\left(
         \int d^4x' \, e^{i (\Lambda_n k) \cdot x'} \, j(x')
     \right) \, .
 \ee
Charge conservation implies that not only the total source current
$\t{J}_\mu$ is transverse,
 \be
 \label{transv}
     k \cdot \t{J}(\vec{k}) = 0 \, ,
 \ee
but that the same is true for the elementary currents whose charge is also
conserved:
 \be
 \label{transv1}
     k' \cdot \t{\jmath}(\vec{k'}) = 0 \, .
 \ee
Due to the randomness of the phases $\phi_n$ (\ref{transv1}) follows
directly from (\ref{transv}) upon inserting (\ref{Jk}) (with
$k'= \Lambda_n k$).

It is well known \cite{GKW79} that a classical current $J^\mu(x)$
generates via Eq.~(\ref{Maxwell}) an asymptotic photon field in a
(normalized) ``coherent'' state
 \be
 \label{coh}
   |J \rangle = e^{-\bar n/2} \exp \left( i \int {d^3k\over (2\pi)^3\, 2\o}
   \sum_{\lambda=1}^2 \t{J}_\mu(\vec{k})\, \eps^\mu(\vec{k},\l) \,
   \hat a^{\dagger}(\vec{k},\l) \right) \vert 0 \rangle \, ,
 \ee
with $\bar n = \int {d^3k\over(2\pi)^3 2\o} \vert \t{J}(\vec{k}) \vert^2$, 
which satisfies
 \be
 \label{aJ}
   \hat a(\vec{k},\l) \, |J \rangle =
   i \, \t{J}_\mu(\vec{k}) \, \eps^\mu(\vec{k},\l) \, |J\rangle \, .
 \ee
Here $\eps^\mu(\vec{k},\l), \, \l{=}1,2,$ are (real and spacelike)
polarization vectors which satisfy the transversality and orthonormality
conditions
 \be
 \label{eps}
   k \cdot \eps(\vec{k},\l)=0\, , \quad
   \eps(\vec{k},\l) \cdot \eps(\vec{k},\l') = g_{\l\l'} \, .
 \ee
According to (\ref{currentx}) the state $\vert J\ra$ depends on the
parameters $\{x_n,p_n,\phi_n;n=1,\dots,N\}$ whose distribution
characterizes the photon emitting source: $\vert J \rangle \equiv
\vert J[N;\{x,p,\phi\}] \rangle$. We take the number $N$ of elementary
currents $j_\mu$ to be distributed
according to a probability distribution $P_N$, the phases $\phi_n$ to be
randomly distributed between 0 and $2\pi$, and the source positions $x_n$
and momenta $p_n$ to be distributed with a classical phase-space density
$\rho(x,p)$, with normalizations
 \be
 \label{norm}
   \sum_{N=0}^\infty P_N =1\, , \quad
   \sum_{N=0}^\infty N\, P_N = \la N \ra \, , \quad
   \int d^4x \, d^4p \, \rho(x,p) = 1\, .
 \ee
The corresponding ensemble average is given by
 \be
 \label{density}
    \la \dots \ra = \sum_{N=0}^\infty P_N
    \int \prod_{n=1}^N d^4x_n \, d^4p_n \, \rho(x_n, p_n)
    \int_0^{2\pi} \frac{d\phi_n}{2\pi} \,
    \la J[N;\{x,p,\phi\}]\vert \dots \vert J[N;\{x,p,\phi\}] \ra \, .
 \ee
Using (\ref{aJ}) it is then straightforward to show that
 \begin{eqnarray}
 \label{P1bare}
   P_1(\vec{k},\l) &=&
   \o \, \eps^\mu(\vec{k},\l) \, \eps^\nu(\vec{k},\l) \,
   \la \t{J}^*_\mu(\vec{k}) \, \t{J}_\nu(\vec{k}) \ra \, ,
 \\
 \label{P2bare}
   P_2(\vec{k}_a,\l_a,\vec{k}_b,\l_b) &=&
   \o_a \, \o_b \, \eps^\mu(\vec{k}_a,\l_a) \, \eps^\nu(\vec{k}_b,\l_b) \,
   \eps^{\bar{\nu}}(\vec{k}_b,\l_b) \, \eps^{\bar{\mu}}(\vec{k}_a,\l_a)
 \nonumber\\
   && \times \la \t{J}^*_\mu(\vec{k}_a) \, \t{J}^*_\nu(\vec{k}_b) \,
      \t{J}_{\bar{\nu}}(\vec{k}_b) \, \t{J}_{\bar{\mu}}(\vec{k}_a) \ra \,.
 \end{eqnarray}
In high energy experiments one averages in practice over the helicities
of the observed photons. Using
 \be
 \label{sum}
    \sum_{\l=1}^2 \eps^\mu(\vec{k},\l)\, \eps^\nu(\vec{k},\l) =
    -g^{\mu\nu} - {k^\mu k^\nu \over (n\cdot k)^2 }
    + {k^\mu n^\nu + n^\mu k^\nu \over n\cdot k} \, ,
 \ee
with an arbitrary timelike unit vector $n^\mu$ with $n^2=1$, one can see
from the transversality condition (\ref{transv}) that the second and third
term on the r.h.s. don't contribute, and we get for the spin-averaged
spectra
 \begin{eqnarray}
 \label{P1sum}
    P_1(\vec{k})
    &=& -\o\, \la \t{J}_\mu^\ast(\vec{k}) \t{J}^\mu(\vec{k}) \ra \, ,
 \\
 \label{P2sum}
    P_2(\vec{k}_a,\vec{k}_b)
    &=& \o_a \, \o_b \, \la \t{J}_\mu^*(\vec{k}_a) \, \t{J}^*_\nu(\vec{k}_b)\,
    \t{J}^\nu(\vec{k}_b) \, \t{J}^\mu(\vec{k}_a) \ra \, .
 \end{eqnarray}

The following steps are completely analogous to those presented in
Ref.~\cite{CH94} (see \cite{H96} for intermediate steps) and will not be
repeated in detail. Performing the average over the random phases leads
to the factorization of the two-particle spectrum (\ref{P2sum}) similar
to the Wick theorem in vacuum and thermal equilibrium systems:
 \be
 \label{P2sum1}
    P_2(\vec{k}_a,\vec{k}_b) =
    \frac{\langle N(N-1) \rangle}{\langle N \rangle^2}
    \Big( P_1(\vec{k}_a)  P_1(\vec{k}_b)
    + \o_a \, \o_b \:
    \la \t{J}_\mu^\ast(\vec{k}_a) \t{J}_\nu(\vec{k}_b) \ra \,
    \la \t{J}^{\nu\ast}(\vec{k}_b) \t{J}^\mu(\vec{k}_a) \ra \Big) \, .
 \ee
Here the identical structure of all elementary sources was required to be 
able to perform the sum over $N$ in (\ref{density}) (see \cite{GKW79,H96} 
for details). The average on the right hand side denotes the remaining 
integrations over the phase-space positions of the elementary sources:
 \be
 \label{aver}
    \la \t{J}_\mu^\ast(\vec{k}_a) \t{J}_\nu(\vec{k}_b) \ra  =
    \la N \ra
    \int d^4x \, e^{-i(k_a-k_b){\cdot}x} \int d^4p\, \rho(x,p)\,
    \left(\bar\Lambda_p\t{\jmath}^*\right)_{\!\mu}\!\!(k_a) \
    \left(\bar\Lambda_p\t{\jmath}\right)_{\!\nu}\!\!(k_b) \, .
 \ee
Defining the Wigner density of the source currents according to
 \be
 \label{wigner}
    S_{\mu\nu}(x,K) = \int d^4y \, e^{-i K{\cdot}y} \,
    \left\langle J^*_\mu(x+{\textstyle{1\over 2}}y)
                  J_\nu(x-{\textstyle\frac{1}{2}}y)
    \right\rangle \, ,
 \ee
the terms on the r.h.s. of (\ref{P2sum1}) can be rewritten as
 \be
 \label{transform}
    \la \t{J}_\mu^\ast(\vec{k}_a) \t{J}_\nu(\vec{k}_b) \ra =
    \t{S}_{\mu\nu}(\vec{q},\vec{K}) \equiv
    \int d^4x \, e^{-i q{\cdot}x}\, S_{\mu\nu}(x,K) \, .
\ee
Here we defined the off-shell vector $K = \frac{1}{2}(k_a + k_b)$ as
the average of the two on-shell photon momenta and $q = k_a - k_b$ as
their difference\footnote{Please note that (although not immediately
apparent from Eq.~(\ref{aver})) the Wigner density in (\ref{transform})
depends only on the average momentum $K$, and not on $k_a$ and $k_b$
separately. This was first derived by Shuryak in his PhD thesis
\cite{S73} and overlooked in Ref.~\cite{RF95}.}. This finally yields
the following form for the two-photon correlation function:
 \be
 \label{corr}
   C(\vec{k}_a,\vec{k}_b) \equiv
   \frac{\langle N \rangle^2}{\langle N(N-1) \rangle} \,
   \frac{P_2(\vec{k}_a,\vec{k}_b)}{P_1(\vec{k}_a)\,P_1(\vec{k}_b)}
   = 1 +
  \frac{\t{S}_{\mu\nu}(\vec{q},\vec{K})\:\t{S}^{\nu\mu}(-\vec{q},\vec{K})}
       {\t{S}_\mu^\mu(\vec{0},\vec{k}_a)\:\t{S}_\mu^\mu(\vec{0},\vec{k}_b)}\;.
 \ee

The difference between this formula and the corresponding one for two-pion
correlations \cite{CH94} is the tensor structure of the Wigner density
(\ref{wigner}) which results from the vector nature of the source currents.
This is where the spin of the photon leaves its traces. To study
spin effects on the 2-photon correlator one must therefore analyze
the tensor structure of the Wigner density $S_{\mu\nu}(x,K)$. To this end
it is useful to factor the elementary source current vectors $j_\mu$ into
their length and a unit vector for their direction. The transversality
condition (\ref{transv1}) allows to decompose the directional unit vector
in the basis spanned by the two polarization vectors:
 \be
 \label{para}
   \left(\bar\Lambda_n \t{\jmath}\right)_\mu\!\!(k) =
   \hat j_n(\vec{k})\, \Bigl(
   \cos\psi_n \, \eps_\mu(\vec{k},1) + \sin\psi_n \, \eps_\mu(\vec{k},2)
   \Bigr)
 \ee
Here $\hat j_n(\vec{k})$ is the length of the vector on the l.h.s., and
$\psi_n$ is an arbitrary angle between 0 and $2\pi$. In the sum over $n$
in Eq.~(\ref{Jk}) $\psi_n$ can take a different value for each term, i.e.
it may be correlated with the momentum $p_n$ of the source $n$.
Eq.~(\ref{para}) is the most general decomposition consistent with
charge conservation. It is, however, more restrictive than the
decomposition suggested in Eq.~(6) of Ref.~\cite{N86} which uses two angles
to parametrize the directional unit vector and thus does not correctly
take into account the constraints from current conservation. As is
obvious from Eqs.~(9) and (10) in that paper \cite{N86}, this oversight
is the origin of the momentum dependent prefactor ${1\over 2} [1 +
\vec{\hat k}_a\cdot \vec{\hat k}_b ]$ in the correlator of Eq.~(8) in
\cite{N86}.

We will now consider the simple case where, except for the constraint
of transversality, the directions of the elementary current vectors
are completely uncorrelated with their momenta and the angle $\psi$ is
a random variable. We believe that this is a reasonable assumption if
the photon emitting source is locally thermalized. The random nature of
$\psi$ is most easily implemented by inserting the Ansatz (\ref{para})
into (\ref{Jk}) and generalizing the ensemble average (\ref{density}) to
include an additional integration over the angles $\psi_n$:
 \begin{eqnarray}
 \label{density1}
    \la \dots \ra &=& \sum_{N=0}^\infty P_N
    \int \prod_{n=1}^N d^4x_n \, d^4p_n \, \rho(x_n, p_n)
 \nonumber\\
    && \times \int_0^{2\pi} \frac{d\phi_n}{2\pi} \frac{d\psi_n}{2\pi} \,
    \la J[N;\{x,p,\phi,\psi\}]\vert \dots
                              \vert J[N;\{x,p,\phi,\psi\}] \ra \, .
 \ee
This procedure gives for the correlation function (\ref{corr}) the simple
result
 \be
 \label{corr2}
   C(\vec{k}_a,\vec{k}_b) = 1 + \frac{1}{2} \,
   \frac{|\t{S}(\vec{q},\vec{K})|^2}
     {\t{S}(\vec{0},\vec{k}_a) \: \t{S}(\vec{0},\vec{k}_b)}
 \ee
where the scalar Wigner density is defined in terms of the {\em moduli}
of the elementary currents in (\ref{para}) (c.f.
Eqs.~(\ref{aver}), (\ref{wigner}) and (\ref{para})):
 \be
   \t{S}(\vec{q},\vec{K}) &=& \int d^4x \, e^{-i q{\cdot}x} S(x,K) \, ,
 \\
   S(x,K) &=& \int d^4y \, e^{-i K{\cdot}y} \, \left\langle
   J^*(x+{\textstyle \frac{1}{2}}y) \, J(x-{\textstyle\frac{1}{2}}y)
   \right\rangle  
   = \la N \ra \int d^4p\; \rho(x,p)\; \hat j_p^*(\vec{k}_a) \; 
     \hat j_p(\vec{k}_b) \, .
   \ee
The factor ${1\over 2}$ in front of the correlation term reflects the
fact that only photons in the same helicity state contribute to the
Bose-Einstein correlations (i.e.~are described with symmetrized
wavefunctions), and that we have summed over final state photon
helicities. Except for this factor, the correlation function has
the same general form as for scalar particles.

We have already explained the origin of the difference between our
result and that of Neuhauser as being due to the neglect of current
conservation in Ref.~\cite{N86}. We would like to close the paper with
a critical discussion of the work by Razumov and Feldmeier \cite{RF95}
who, using a different derivation, found similar spurious terms in the
tensor structure of the photon emission function (\ref{wigner}).
These authors used the Ansatz (Eq.~(16) of Ref.~\cite{RF95})
 \be
 \label{parRF}
    \la \t{J}^{\ast\mu}(\vec{k}_a) \, \t{J}^\nu(\vec{k}_b) \ra
    \stackrel{\rm Ansatz}{=}
    \int d^4x \, \e^{-i q{\cdot}x} Q^{\mu\nu}(k_a,k_b|x)
    \,\omega(k_a,k_b|x)\, ,
 \ee
which (as is also clear from the text of their paper) should be compared
with eq.~(\ref{transform}) above. They then performed two (in our opinion)
erroneous manipulations: using the transversality of the correlator
 \be
 \label{trans}
    k_a^\mu \, \la \t{J}_\mu^\ast(\vec{k}_a) \, \t{J}_\nu(\vec{k}_b) \ra
    = \la \t{J}_\mu^\ast(\vec{k}_a) \, \t{J}_\nu(\vec{k}_b) \ra
    \, k_b^\nu = 0
 \ee
they argued that the tensor $Q^{\mu\nu}$ in (\ref{parRF}) should also be
transverse to $k_a$ and $k_b$. This, however, does not follow: writing
$k_{a,b} = K \pm q/2$ and realizing (see (\ref{transform})) that
$Q(k_a,k_b\vert x)$ is actually only a function of $K=(k_a+k_b)/2$, one
realizes that the transversality condition for $Q$ is actually much more
complicated and involves derivates of $Q$ with respect to $x$. Furthermore,
in Eq.~(20) of Ref.~\cite{RF95} the tensor structure of $Q$ is parametrized
in terms of $k_a$ and $k_b$ separately instead of only in terms of
$K$, leading to spurious terms in the tensor structure which eventually
result in their spurious momentum-dependent prefactor in the second term of
the correlation function (\ref{corr2}). For a hydrodynamicaly expanding
source their decomposition of the $Q$-tensor also involves the
hydrodynamical 4-velocity. From Eqs.~(\ref{aver}), (\ref{transform})
it is clear that in a hydrodynamical parametrization of the emission
function the hydrodynamical flow velocity enters through the classical
phase-space density $\rho(x,p)$ of the elementary source currents,
which (in Boltzmann approximation) will have the typical local equilibrium
form $\rho(x,p) \sim \exp(-p\cdot u(x)/T(x))$. It is not directly associated
with the elementary currents $\t{\jmath}_\mu$ which generate
the tensor structure $Q^{\mu\nu}(K,x)$ of the emission function.

In summary, we have rederived the general form for the two-photon
correlation function for chaotic sources using the covariant current
formalism. We found additional constraints on the tensor structure
of the photon emission function resulting from conservation of the
electromagnetic current which were previously overlooked. For
thermalized sources with random orientation of the elementary
photon-emitting currents we found that the only effect of photon spin
on the correlation function is a reduction of the correlation strength
by a factor ${1\over 2}$ resulting from spin averaging in the detector.
Similarly simple results are expected to hold for quantum statistical
correlations between identical fermions, justifying the generally adopted
procedure in the analysis of $pp$ and $nn$ correlations.

\acknowledgments
We are indebted to S. Leupold and B. Tom\'a\v{s}ik for valuable
comments. This work was supported by BMBF, DFG, and GSI.


\end{document}